\documentclass[a4paper]{article}
\usepackage{afterpage,latexsym,epsfig,float,tabularx, amsmath,cite}

\newcolumntype{Y}{>{\centering\arraybackslash}X}
\newcommand{\bra}{\langle}
\newcommand{\ket}{\rangle}
 
\newcommand{\Hrule}{\rule{\linewidth}{1mm}}

\restylefloat{figure}

\begin{document}

\newpage

\thispagestyle{empty}

\vspace*{\stretch{1}}
\noindent
\Hrule
\begin{center}
  \huge {\bf Introduction to Monte Carlo methods for an \\ Ising Model of a Ferromagnet} \\[8mm]
  \large 
  \begin{quotation}
    {\itshape `If God has made the world a perfect
      mechanism, He has at least conceded so much to our imperfect
      intellects that in order to predict little parts of it, we need
      not solve innumerable differential equations, but can use dice
      with fair success.' }
      \begin{flushleft}
        Max Born
      \end{flushleft}
  \end{quotation}
\end{center}
\Hrule
\vspace*{\stretch{2}}
\begin{flushright}
  \Large{Jacques Kotze}
\end{flushright}

\newpage
\setcounter{page}{1}

\tableofcontents
\vspace{1.5cm}
\noindent
{\large {\bf Acknowledgements}}
\vspace{0.5cm}

Thank you to Prof B.A. Bassett for encouraging me to submit this report and my office colleagues in Room 402 for their assistance in reviewing it. A special thanks must be extended to Dr A.J. van Biljon who was always
available for invaluable discussion, advice and assistance. Dr
L. Anton and Dr~K.~M\"uller-Nederbok also helped in sharing their expertise 
with regards this report. Lastly I would like to thank Prof
H.B. Geyer for his patience and constructive criticism.
\newpage

 
\section{Theory}

\subsection{Introduction}

This discussion serves as an introduction to the use of Monte Carlo simulations as a useful way to evaluate the observables of a ferromagnet. Key background is given about the relevance and effectiveness of this stochastic approach and in particular the applicability of the Metropolis-Hastings algorithm. Importantly the potentially devastating effects of spontaneous magnetization are highlighted and a means to avert this is examined. 

An Ising model is introduced and used to investigate the properties of a two dimensional ferromagnet with respect to its magnetization and energy at varying temperatures. The observables are calculated and a phase transition at a critical temperature is also illustrated and evaluated. Lastly a finite size scaling analysis is underatken to determine the critical exponents and the Curie temperature is calculated using a ratio of cumulants with differing lattice sizes. The results obtained from the simulation are compared to exact calculations to endorse the validity of this numerical process. A copy of the code used, written in C++, is enclosed and is freely available for use and modification under the General Public License.
%

\subsection{Background}\label{sec:back}

In most ordinary materials the associated magnetic dipoles of the atoms have a
random orientation. In effect this non-specific distribution results
in no overall macroscopic magnetic moment. However in certain cases,
such as iron, a magnetic moment is produced as a result of a preferred
alignment of the atomic spins. 

This phenomenon is based on two fundamental principles, namely \emph{energy
minimization} and \emph{entropy maximization}. These are competing 
principles and are important in moderating the overall effect.
Temperature is the mediator between these opposing elements
and ultimately determines which will be more dominant.

\begin{figure}[hbt]
  \centering
  \epsfig{file=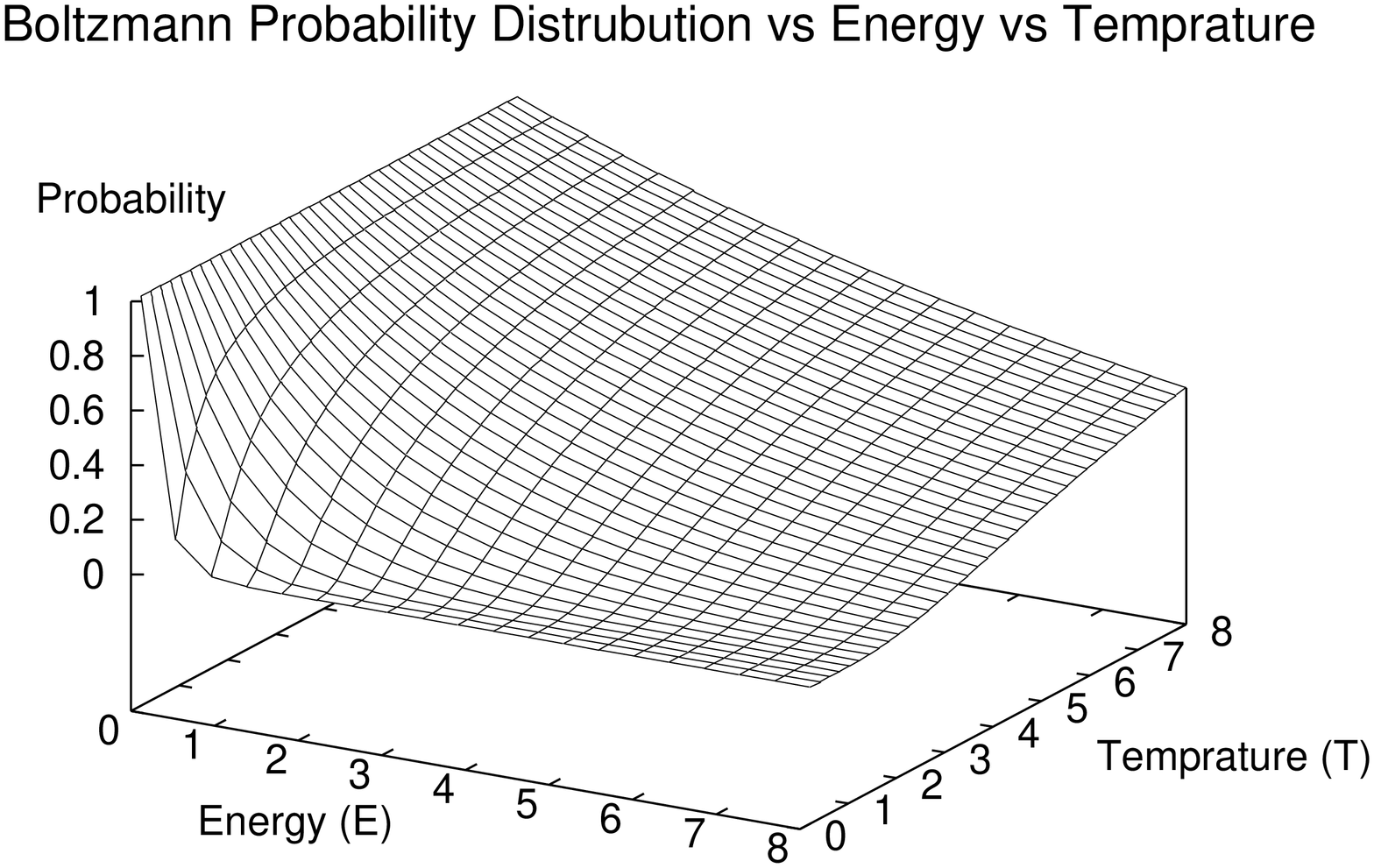,height=9cm,width=5.5cm,angle=270,clip=,bbllx=10pt,bblly=10pt,bburx=510pt,bbury=820pt}
  \caption{This figure shows the Boltzmann probability distribution as a landscape for varying Energy ($E$) and Temperature ($T$).}\label{fig:prob}
\end{figure}

The relative importance of the energy minimization and entropy maximization is
governed in nature by a specific probability

\begin{equation}
P(\alpha)=\exp{\frac{-E(\alpha)}{kT}}\label{eq:boltzmann}.
\end{equation}
which is illustrated in Figure \ref{fig:prob} and is known as the Gibbs
distribution. 

\subsection{Model}

A key ingredient in the understanding of this theory is the spin and
associated magnetic moment. Knowing that spin is a quantum mechanical
phenomenon it is easy to anticipate that a complete and thorough
exposition of the problem would require quantum rules of spin and angular 
momentum. These factors prove to be unnecessary complications.

We thus introduce a model to attain
 useful results. The idea central to a
model is to simplify the complexity of the problem to such a degree 
that it
is mathematically tractable to deal with while retaining the
essential physics of the system. The Ising Model does this very
effectively and even allows for a good conceptual understanding.

\begin{figure}[htb]
  \centering
  \epsfig{file=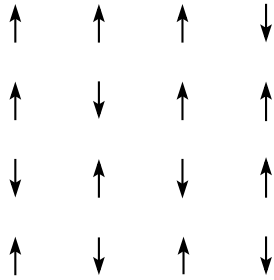,height=4cm,width=4cm,angle=0,clip=,bbllx=0pt,bblly=0pt,bburx=260pt,bbury=280pt}
  \caption{Two Dimensional lattice illustration of an Ising Model. The up and down arrows represent a postive and a negative spin respectively.  }\label{fig:isinglat}
\end{figure}
\begin{figure}[htb]
  \centering
  \epsfig{file=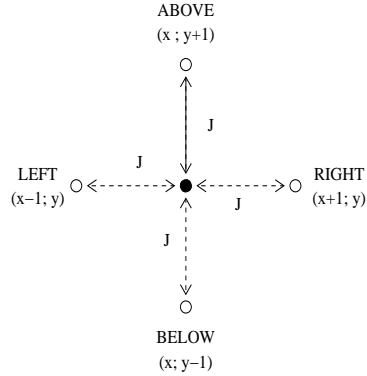,height=5cm,width=5cm,angle=0,clip=,bbllx=0pt,bblly=10pt,bburx=310pt,bbury=290pt}
  \caption{Nearest neighbour coupling. The dark dot, at position (x,y), is being interacted upon by its nearest neighbours which are one lattice spacing away from it.}\label{fig:Jnn}
\end{figure}

The Ising Model considers the problem in two 
dimensions\footnote{It may also be considered in three dimensions but
  it should be noted that a one dimensional representation does not
  display phase transition.} and places dipole spins at regular
lattice points while restricting their spin axis to
be either up (+y) or down (-y). The lattice configuration is square
with dimensions $L$ and the total number of spins equal to $N=L\times L$. In its simplest form the interaction
range amongst the dipoles is restricted to immediately adjacent
sites (nearest neighbours). This produces a Hamiltonian for a specific
spin site, $i$, of the form 

\begin{equation}
H_i=-J \sum_{j_{nn}} s_i s_j\label{eq:ham}
\end{equation}
where the sum $j_{nn}$ runs over the nearest neighbours of $i$.
The coupling constant between nearest neighbours is represented by 
$J$ while the $s_i$  and $s_j$ are the respective nearest neighbour
spins.  The nature of the interaction in the model is all contained in the
sign of the interaction coupling 
constant $J$. If $J$ is positive it would mean that the material has a
ferromagnetic nature (parallel alignment) while a negative sign would
imply that the material is antiferromagentic (favours anti-parallel
alignment). $J$ will be taken to be +1 in our discussion and the values for spins
will be +1 for spin up and -1 for spin down. A further simplification
is made in that $J/k_b$ is taken to be unity. The relative positioning
of nearest neighbours of spins is shown in Figure \ref{fig:Jnn} with the
darker dot being interacted on by its surrounding neighbours. 

To maximize the interaction of the spins at the edges of the lattice
they are made to interact with the spins at the geometric opposite edges
of the lattice. This is referred to as periodic boundary 
condition (pbc) and can be visualized better if we consider the 2d
lattice being folded into a 3d torus with spins being on the surface
of this topological structure. 

\begin{figure}[htb]
  \centering
  \epsfig{file=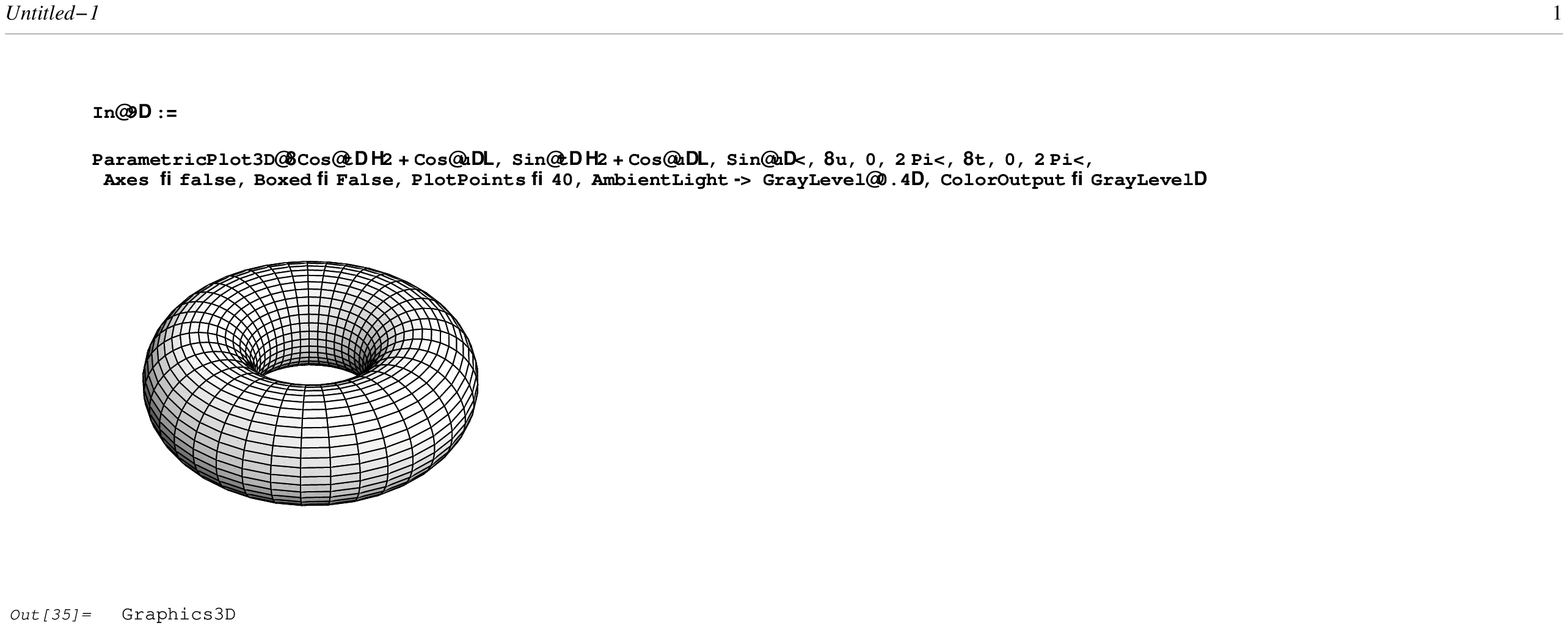,height=6cm,width=5cm,angle=270,clip=,bbllx=140pt,bblly=120pt,bburx=280pt,bbury=280pt}
  \caption{An illustration of a three dimensional torus which is repreentative of a two dimensional lattice with periodic boundary conditions.}\label{fig:pbc}
\end{figure}

\subsection{Computational Problems}

With the help of an Ising Model we can proceed with the
anticipation of achieving solutions for observables.
If the energy of each possible state of the system is specified, then
the Boltzmann distribution function, equation \eqref{eq:boltzmann}, gives the probability
for the system to be in each possible state (at a given temperature)
and therefore macroscopic quantities of interest can be calculated by
doing probability summing. This can be illustrated by using
magnetization and energy as examples. For any 
fixed state, $\alpha$, the magnetization is proportional to the `excess'
number of spins pointing up or down while the energy is given by the
Hamiltonian \eqref{eq:ham}.

\begin{equation}
  \ M(\alpha)=N_{up}(\alpha)-N_{down}(\alpha)
\end{equation}
The expected value for $M$ is given by 

\begin{equation}
  \label{eq:prob_sum_M}
  \bra M \ket=\sum_\alpha M(\alpha)P(\alpha),
\end{equation}
and the expected value for $E$ is given by
\begin{equation}
  \label{eq:prob_sum_E}
  \langle E \rangle=\sum_\alpha E(\alpha)P(\alpha).
\end{equation}
These calculation pose a drastic problem from a practical
perspective. Considering we have two spin orientations (up \& down) and 
there are $N$ spins which implies that there are $2^N$ different
states. As $N$ becomes large it is evident that computations become a
daunting task if calculated in this manner. 

It may seem a natural suggestion to use a computer simulation to
do these calculations but by examining equations \eqref{eq:prob_sum_M} 
and \eqref{eq:prob_sum_E}
more closely it becomes apparent that using this procedure would waste as much
computing effort on calculating an improbable result as it does on a very 
probable result. Thus a better numerical alternative would be to use a
simulation to generate data over the `representative states'. These
representative states constitute the appropriate proportions of different states
\footnote{The frequency with which some class of events is encountered 
  in the representative sum must be the same as the actual probability
  for that class.}. This is a form of biased sampling which
essentially boils down to satisfying the following condition  
\vspace{1mm}
\begin{center}
  {\bf GENERATED FREQUENCY}{\bf $\equiv$}{\bf ACTUAL PROBABILITY}.
  {\bf \hspace{1cm}\vspace{1mm}(computer)\hspace{3.5cm}(theory)}
\end{center}We now examine, in a more formal setting, how to accomplish 
this objective.

\subsection{Sampling and Averaging}

\begin{figure}[htb]
  \centering
  \epsfig{file=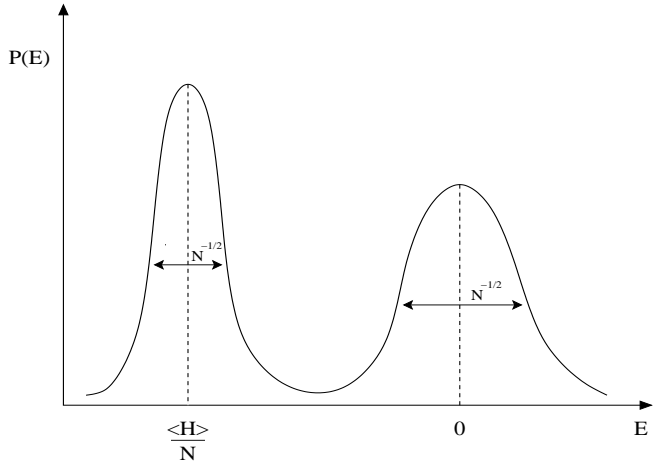,height=6.5cm,width=9cm,angle=0,clip=,bbllx=10pt,bblly=10pt,bburx=550pt,bbury=450pt}
  \caption{Example of a simple sampling producing a Gaussian
    distribution, centered around zero, while the crucial data is
    peaked outside the sampling region.}\label{fig:simp_sampling}  
\end{figure}

The thermal average for an observable $A(x)$ is defined in the
canonical ensemble   

\begin{equation}
  \bra A(x) \ket_T = \frac{1}{Z} \int e^{-\beta H(x)} A(x)
  dx \label{eq:ensemble_average} 
\end{equation}
\noindent where $x$ is a vector in the phase space and $\beta=1/k_b T$. The Partition
function, $Z$, is given by

\begin{displaymath}
  Z = \int e^{-\beta H(x)} dx
\end{displaymath}
\noindent while the normalized Boltzmann factor is 

\begin{equation}
  P(x) = \frac{1}{Z} e^{-\beta H(x)}. \label{eq:normalized_boltz}
\end{equation}
This probability gives the actual statistical weight with which the
configuration $x$ occurs in the thermal equilibrium. We now want to
consider the discrete case of the formal definitions above. If we are
to consider a finite portion of the phase space it would produces an
average of the form   

\begin{equation}
  \bra A(x) \ket =\frac{\sum_{l=1}^{M} e^{-\beta H(x_l)}
    A(x_l)}{\sum_{l=1}^{M} e^{-\beta H(x_l)}}.\label{eq:discreet_avg}
\end{equation}
If we were to take $M\rightarrow \infty$ in equation
\eqref{eq:discreet_avg} it would reduce to equation
\eqref{eq:ensemble_average}. The problem with taking a simple sample
of this nature in the phase space is that it would not guarantee that the probability
distribution is peaked in the region under consideration (not
representative). Figure \ref{fig:simp_sampling} illustrates this problem.  

It thus makes sense to attempt a smarter sampling technique to include 
the important areas in the phase space. We want a process that selects 
points, $x_l$, with an associated probability, $P(x_l)$ in the phase
space. Estimating the thermal average now for a chosen set, ${x_l}$,
reduces equation \eqref{eq:discreet_avg} to 

\begin{equation}
  \bra A(x) \ket =\frac{\sum_{l=1}^{M} e^{-H(x_l)\beta}
    A(x_l)/P(x_l)}{\sum_{l=1}^{M} e^{-H(x_l)\beta}/P(x_l)}.\label{eq:discreet_avg_imp_samp}
\end{equation} 
The most sensible choice for $P(x_l)$ is $P(x_l) \propto
e^{-H(x_l)\beta} \label{eq:imp_samp_prob}$. This construct produces the
simple arithmetic average for \eqref{eq:discreet_avg_imp_samp} by
canceling out the Boltzmann factors, thus

\begin{equation}
  \bra A(x) \ket =\frac{1}{M}{\sum_{l=1}^{M}
    A(x_l)}.\label{eq:arithmetic_avg}
\end{equation}  
If we stop to reflect on what we are trying to achieve at this point,
we discover that we are attempting to reduce a probability distribution at
equilibrium of the infinite phase space to a representative
distribution with a finite set of points from the phase space,
${x_l}$. The question is how do we generate this distribution.

Metropolis et al. advanced conventional thinking at the time by
introducing the idea of using a Markov process of successive states
${x_l}$ where each state $x_{l+1}$ is constructed from a previous
state $x_l$ via a suitable transition probability $W(x_l\rightarrow
x_{l+1})$. In order to implement this idea successfully a detailed
balance equation has to be imposed,

\begin{equation}
  P_{eq}(x_l)W(x_l\rightarrow x_{l'}) =P_{eq}(x_{l'})W(x_{l'}\rightarrow
  x_{l}). \label{eq:detailed_balance}
\end{equation}  
This equation simply prescribes that at equilibrium there is an equal
probability for $x_l \rightarrow x_{l'}$ and $x_{l'} \rightarrow
x_{l}$. If we now take the ratio of the transition probabilities it
becomes evident that a move $x_l \rightarrow x_{l'}$ and the inverse
move $x_{l'} \rightarrow x_{l}$ is only dependent on the energy change 
$\delta H= H(x_{l'})-H(x_l)$.

\begin{equation}
  \label{eq:ratio_transition_prob}
  \frac{W(x_l \rightarrow x_{l'})}{W(x_{l'} \rightarrow x_{l})} =
  e^{-\delta H \beta}
\end{equation}
This doesn't however help to specify $W(x_l \rightarrow x_{l'})$
uniquely. In order to do this we introduce

\begin{equation}
  \label{eq:complete_trans_prob}
  W(x_l \rightarrow x_{l'})=
  \begin{cases}
    e^{-\delta H \beta} & \text{if $\delta H < 0$ ,}\\
    & \\
    1 & \text{otherwise ,}
  \end{cases}
\end{equation}
It can be shown that using this transition probability
$W(x_l\rightarrow x_{l+1})$ the distribution $P(x_l)$ of states
generated by the Markov process tend to the equilibrium
distribution as $M \rightarrow \infty$. Thus the construct holds and
approximates the theory with an increasing degree of accuracy as we
consider a larger set of points, $\{ x_l \}$, in the phase space.

The changes in the probability distribution over time are governed by
the Markovian Master equation

\begin{equation}
  \label{eq:markovian_master}
  \frac{dP(x,t)}{dt} = - \sum_{x'} W(x_l \rightarrow
  x_{l'}) + \sum_{x'} W(x_{l'} \rightarrow x_{l}).
\end{equation}
In the thermal limit where $P(x_l) = P_{eq}(x_l)$ the detailed balance
equation, \eqref{eq:detailed_balance}, which was imposed, comes into
play resulting in $dP(x,t)/dt = 0$ as we would expect. Furthermore since 
we are considering a finite system it is logical to conclude that the
system is ergodic\footnote{Attribute of a behavior that involves only equilibrium states and whose transition probabilities either are unvarying or follow a definite cycle. In statistics, ergodicity is called stationarity and tested by comparing the transition probabilities or different parts of a longer sequence of events.}. 

\begin{equation}
  \bra A(t) \ket = \frac{1}{t_M} \int A(t) dt\label{eq:temp_average} 
\end{equation}
This relation reduces in a similar fashion to the arithmetic average previously discussed
if we consider the number of Monte Carlo steps (mcs) as  a units of
`time'. We thus are confronted by the question of whether the system
is ergodic in order for the time averaging to be equivalent to the
canonical ensemble average. This condition is thus forced upon the
system if we consider the mcs as a measure of time.

\begin{equation}
  \bra A(t) \ket =\frac{1}{M}\sum_{t=1}^{M}A(x(t)) 
  \label{eq:discreet_temp_avg}
\end{equation}
Thus the Metropolis sampling can be interpreted as time averaging along
a stochastic trajectory in phase space, controlled by the master equation
\eqref{eq:markovian_master} of the system. 

\subsection{Monte Carlo Method}

The fact that we can make a successful stochastic interpretation of
the sampling procedure 
proves to be pivotal in allowing us to introduce the Monte Carlo method
as a technique to solve for the observables of the Ising Model.

A Monte Carlo calculation is defined, fundamentally, as explicitly
using random variates and a stochastic process is characterized by a random 
process that develops with time. The Monte Carlo method thus lends
itself very naturally to simulating systems in which stochastic
processes occur. From what we have established with
regards to the sampling being analogous to a time averaging along a
stochastic trajectory in the phase space it is possible to simulated
this process by using the Monte Carlo method. The algorithm used is
design around the principle of the Metropolis sampling.

From an implementation point of view the issue of the random number lies
at the heart of this process and its success depends on the fact that
the generated number is truly random. 
`Numerical Recipes in C', \cite{NumRecipe}, deals with this issue
extensively and the random number generator `Ran1.c' from this text
was used in the proceeding simulation. This association with
randomness is also the origin of the name of the simulation since the
glamorous location of Monte Carlo is synonymous with luck and chance. 

\subsection{Calculation of observables}

The observables of particular interest are $\bra E \ket,\ \bra E^2 \ket, 
\bra M \ket, \bra |M| \ket$ and $\bra M^2 \ket$. These are calculated in
the following way,

\begin{equation}
  \label{eq:magnetization}
  \bra M \ket = \frac{1}{N}\sum_{\alpha}^{N} M(\alpha)
\end{equation}similarly $\bra |M| \ket$ and $\bra M^2 \ket$ are calculated using the
above equation.

To calculate energy we use the Hamiltonian given in equation
\eqref{eq:ham},

\begin{equation}
  \label{eq:energy}  
  \bra E \ket=\frac{1}{2}\langle \ \sum_{i}^{N} H_i\ \rangle =\frac{1}{2}\langle \ -J
  \sum_{i}^{N}\sum_{j_{nn}} s_i s_j\ \rangle 
\end{equation}the factor of a half is introduced in order to account
for the spins being counted twice. Equation \eqref{eq:energy} is used
in a similar way to determine $\bra E^2 \ket$. 

At the Curie temperature we
expect a marked fluctuation in these quantities. A good candidate to
illustrate this fluctuation would be the variance $(\Delta A)^2 =~\bra
A^2 \ket - \bra A \ket^2$. This leads us to the logical conclusion of
calculating the heat capacity, $C$, and the susceptibility, $\chi$.

\begin{equation}
  \label{eq:heat_capacity}
  C =  \frac{\partial E}{\partial T} = \frac{(\Delta E)^2}{k_b T} =
  \frac{\bra E^2 \ket - \bra E \ket^2}{k_b T^2} 
\end{equation}

\begin{equation}
  \label{eq:susceptibility}
  \chi = \frac{\partial M}{\partial T} = \frac{(\Delta M)^2}{k_b T} =
  \frac{\bra M^2 \ket - \bra M \ket^2}{k_b T} 
\end{equation}

A cumulant is also calculated. This will be used to ultimately determine the
Curie temperature.  
\begin{equation}
  \label{eq:cum1}
  U_L=1-\frac{\langle M^4 \rangle_L}{3\langle M^2 \rangle_L}
\end{equation}

\subsection{Metropolis Algorithm}
\begin{figure}[htb]
  \centering 
  \epsfig{file=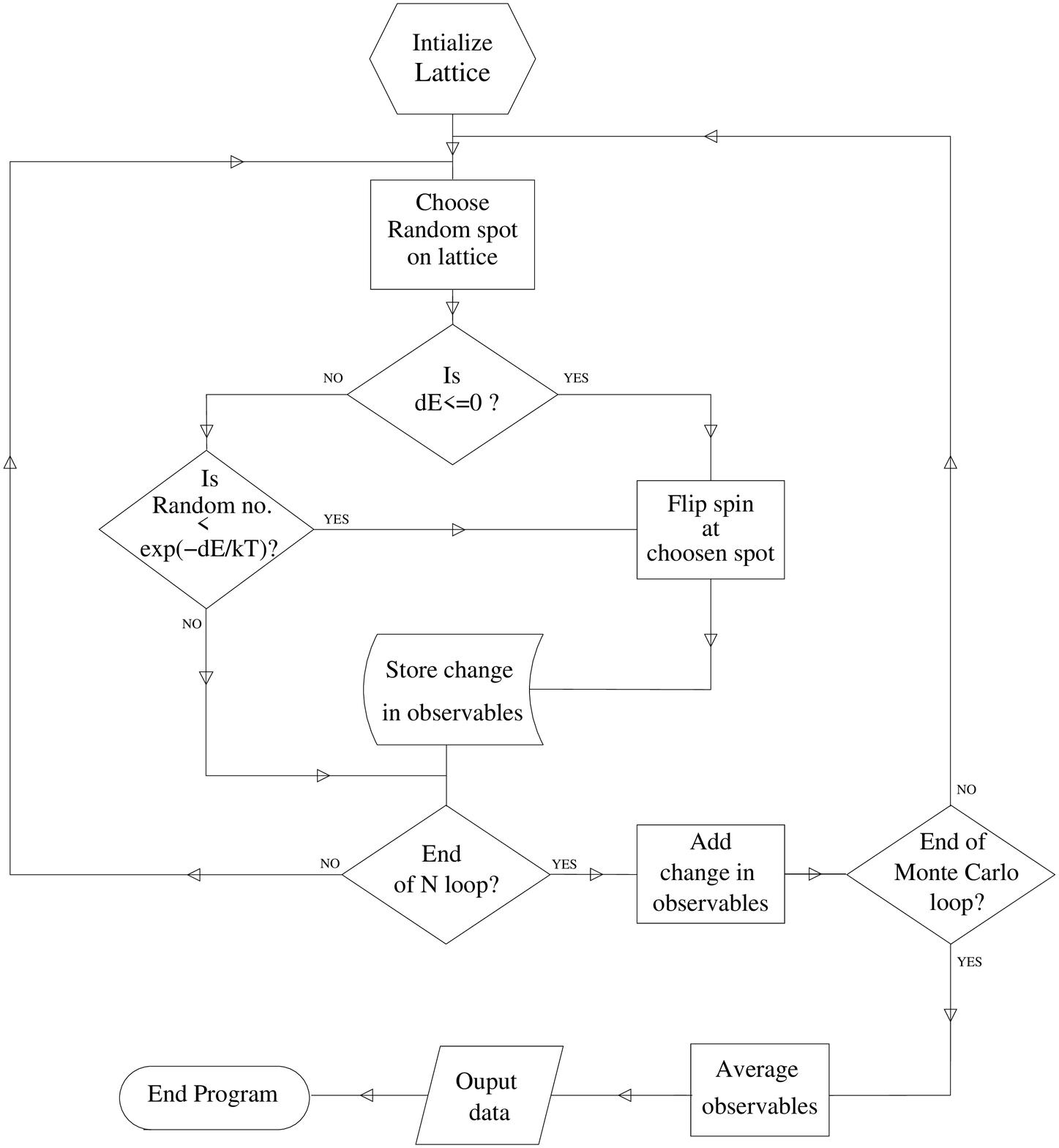,height=12cm}
\caption{Metropolis Flowchart}\label{fig:flowchart}
\end{figure}

The algorithm implemented for this simulation is the Metropolis
Algorithm \cite{Metropolis:53}. The steps executed in the program are best summarized
in a flowchart. From the flowchart, Figure \ref{fig:flowchart}, it is possible to attain a
better conceptual feel for what the algorithm is attempting to
achieve. 

\begin{itemize}
\item{In the first step the lattice is INITIALIZED to a
    starting configuration. This may either be a homogeneous or random 
    configuration. A random configuration has a benefit in that it
    uses less 
    computing time to reach a equilibrated configuration with the
    associated heat bath.}
\item{In the following PROCESS the random number generator is used to
    select a position on the lattice by producing a uniformly
    generated number between 1 and $N$.}
\item{A DECISION is then made whether the change in energy of
    flipping that particular spin selected is lower than zero. This is 
    in accordance with the principle of energy minimization.}
  \begin{itemize}
  \item{If the change in energy is lower than zero then a PROCESS is
      invoked to flip the spin at the selected site and the associated 
      change in
      the observables that want to be monitored are stored.}
  \item{If the change in energy is higher than zero then a DECISION
      has to be used to establish if the spin is going to be flipped,
      regardless of the higher energy consideration. A random number
      is generated between 0 and 1 and then weighed against the
      Boltzmann Probability factor. If the random number is less than
      the associated probability, $e^{-\delta \beta H}$, then the spin
      is flipped (This would allow for the spin to be flipped as a
      result of energy absorbed from the heat bath, as in keeping
      with the principle of entropy maximization) else it is left
      unchanged in its original configuration.}  
  \end{itemize} 
\item{The above steps are repeated N times and checked at this point in a DECISION
    to determine if the loop is completed. The steps referred to here
    do not include the initialization which is only required once in
    the beginning of the algorithm.} 
\item{Once the N steps are completed a PROCESS is used to add all the
    progressive changes in 
    the lattice configuration together with the original
    configuration in order to produce a new lattice configuration.}  
\item{All these steps are, in turn, contained within a Monte
    Carlo loop. A DECISION is used to see if these steps are
    completed.} 
\item{Once the Monte Carlo loop is completed the program is left with, 
    what amounts to, the sum of all the generated lattices within the
    N loops. A PROCESS is thus employed to average the accumulated
    change in observables over the number of spins and the number of
    Monte Carlo steps.}
\item{Lastly this data can be OUTPUT to a file or plot.} 
\end{itemize}

This run through the algorithm produces a set of observables for a
specific temperature. Considering that we are interested in
seeing a phase transition with respect to temperature 
we need to contain this procedure within a temperature loop in order to 
produce these observables for a range of temperatures. 

The program that was implemented for this discussion started at a
temperature of $T=5$ and progressively stepped down in temperature to
$T=~0.5$ with intervals of $\delta T=0.1$. The different lattice
sizes considered where $2\times 2$, $\ 4\times 4$, $\ 8\times 8 \ $and$ 
\ 16\times 16$. A million Monte Carlo steps (mcs) per spin where used in
order to ensure a large sample of data to average over. The physical
variables calculated were $\bra E \ket,\ \bra E^2 \ket, \ \bra |M|
\ket \ $and$ \ \bra M^2 \ket$. 

A problem that occurs after the initialization is that the
configuration will, more than likely, not be in equilibrium with the
heat bath and it will take a few Monte Carlo steps to reach a
equilibrated state. The results produced during this period are
called transient and aren't of interest. We thus have to make
provision for the program to 
disregard them. This is achieved by doing a run of a thousand Monte
Carlo steps preceding the data collection of any statistics for a
given temperature in order to ensure that it has reached a
equilibrated state. This realization is only significant for the
initial configuration and the problem is avoided by the program in the
future by using small temperature steps and the 
configuration of the lattice at the previous temperature. A very small 
number of mcs are thus required for the system to stabilize its
configuration to the new temperature. 



\section{Results}

\subsection{Energy Results}\label{sec:energy-results}

\begin{figure}[htb]
  \centering 
  \epsfig{file=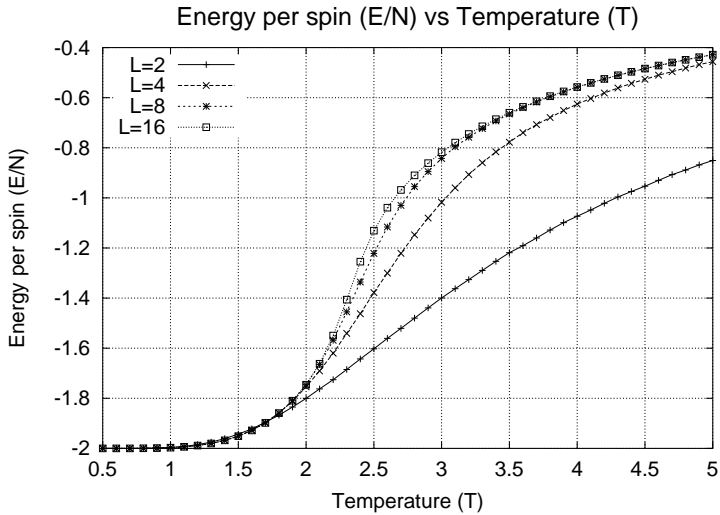,width=7cm,angle=270}
  \caption[hjsagh]{This plot shows the differing results of the Energy 
    for varying lattice
    sizes, $L\times L$.}\label{fig:EvsT}
\end{figure}
In Figure \ref{fig:EvsT} the energy per spin as a function of temperature can be
seen. The curve of the graph becomes more pronounced as the lattice
size increases but there isn't a marked difference between the $L=8$ and
$L=16$ lattices. The steep gradient in the larger lattices points
towards a possible phase transition but isn't
clearly illustrated. The energy per spin for higher temperatures is
relatively high
which is in keeping with our expectation of having a random
configuration while it stabilizes to a $E/N=-2J=-2$ at low
temperatures. This indicates that the spins are all aligned in
parallel. 

\begin{figure}[H]
  \centering
  \epsfig{file=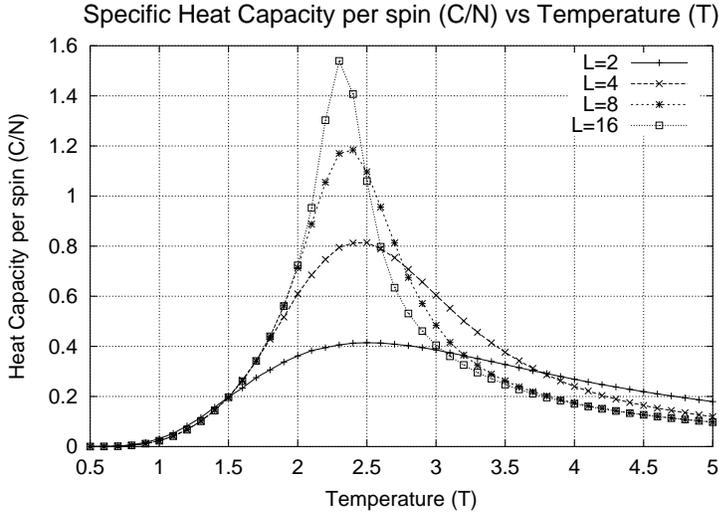,width=7cm,angle=270}
  \caption[sffs]{This plot shows the differing results of Specific
    Heat Capacity for varying lattice
    sizes, $L\times L$.}\label{fig:CvsT}
\end{figure}
In Figure \ref{fig:CvsT} the specific heat capacity per spin is shown as a
function of temperature. We concluded previously that a divergence
would occur at a phase transition and thus should be looking for such
a divergence on the graph. It is however clear that there is no such
divergence but merely a progressive steepening of the peak as the
lattice size increases. The point at which the plot is peaked should be
noted as a possible point of divergence. The reason for not explicitly 
finding a divergence will be discussed in Section \ref{sec:finite-size-scaling}.

\subsection{Magnetization Results}\label{sec:magn-results}

\begin{figure}[htb]
  \centering
\epsfig{file=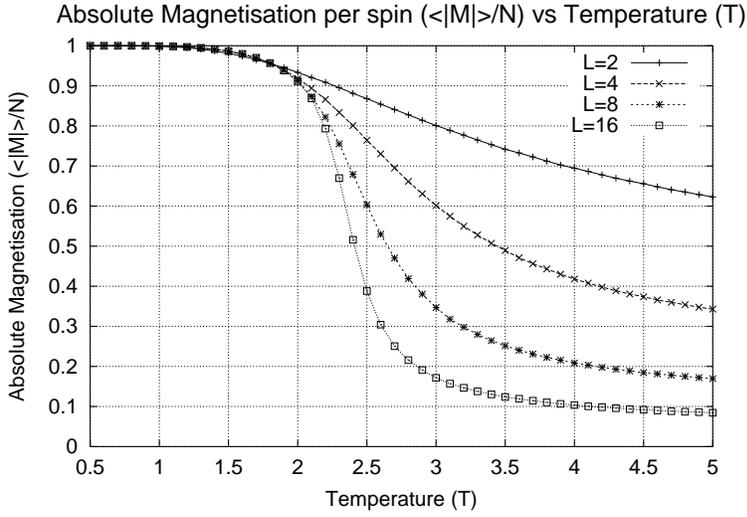,width=7cm,angle=270}
  \caption[hjsgd]{This plot shows the differing results of the
    Magnetization for varying lattice
    sizes, $L\times L$.} \label{fig:MvsT}
\end{figure}
Figure \ref{fig:MvsT} of the magnetization results shows very beautifully
that the shape of the gradient becomes more distinct as the lattice
size is increased. Furthermore, as opposed to Figure \ref{fig:EvsT},
there is a far more apparent difference that the larger lattices produce
in the curves and this illustrates a more apparent continuous phase
transition. The behaviour of the magnetization at high and low
temperature are as the theory prescribes (random to stable parallel
aligned configuration).

At this juncture it is prudent to point out that the susceptibility cannot
be calculated using the ordinary technique in equation
\eqref{eq:susceptibility} given in the discussion on the calculation of
observables. The reason is focused around a subtle fact that has
drastic implications. To comprehend the problem
at work we have to consider one of the constraints of our model,
namely the finite nature of our lattice. This manifests in the fact
that spontaneous magnetization can occur for a finite sized lattice. In
this instance the effect is of particular interest below the critical
temperature.

\begin{figure}[htb]
  \centering
  \epsfig{file=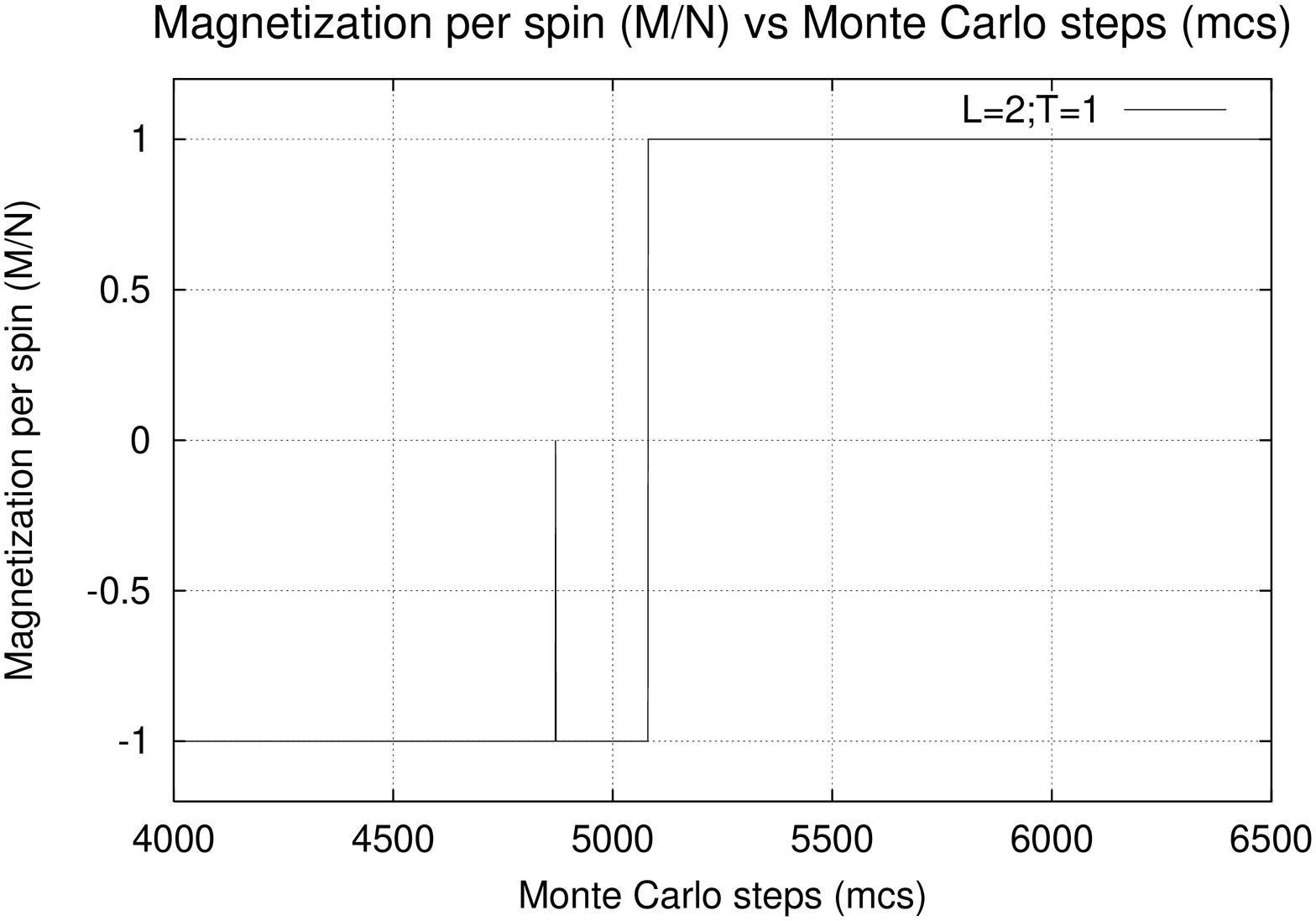,height=10.5cm,width=7cm,angle=270}
  \caption{This plot shows a spontaneous flip in magnetization for a
    $2\times 2$ lattice at $T=1$.} \label{fig:sponmag}
\end{figure}
This can be illustrated by considering the following example of
collected data in Figure \ref{fig:sponmag}.
This data is taken at a temperature that is considerably less than the 
Curie temperature and we would thus expect it to have a stable nature
and yet it clearly displays a fluctuation that is uncharacteristic, resulting in a complete
flip of the magnetization. It has already
been highlighted that because we are dealing with a limited lattice
size there is finite probability for this kind of behaviour to take
place. This probability is directly proportional to the number of mcs used and
inversely proportional to the lattice size, this is compounded by the preiodic boundary conditions used. 

\begin{figure}[htb]
  \centering
  \epsfig{file=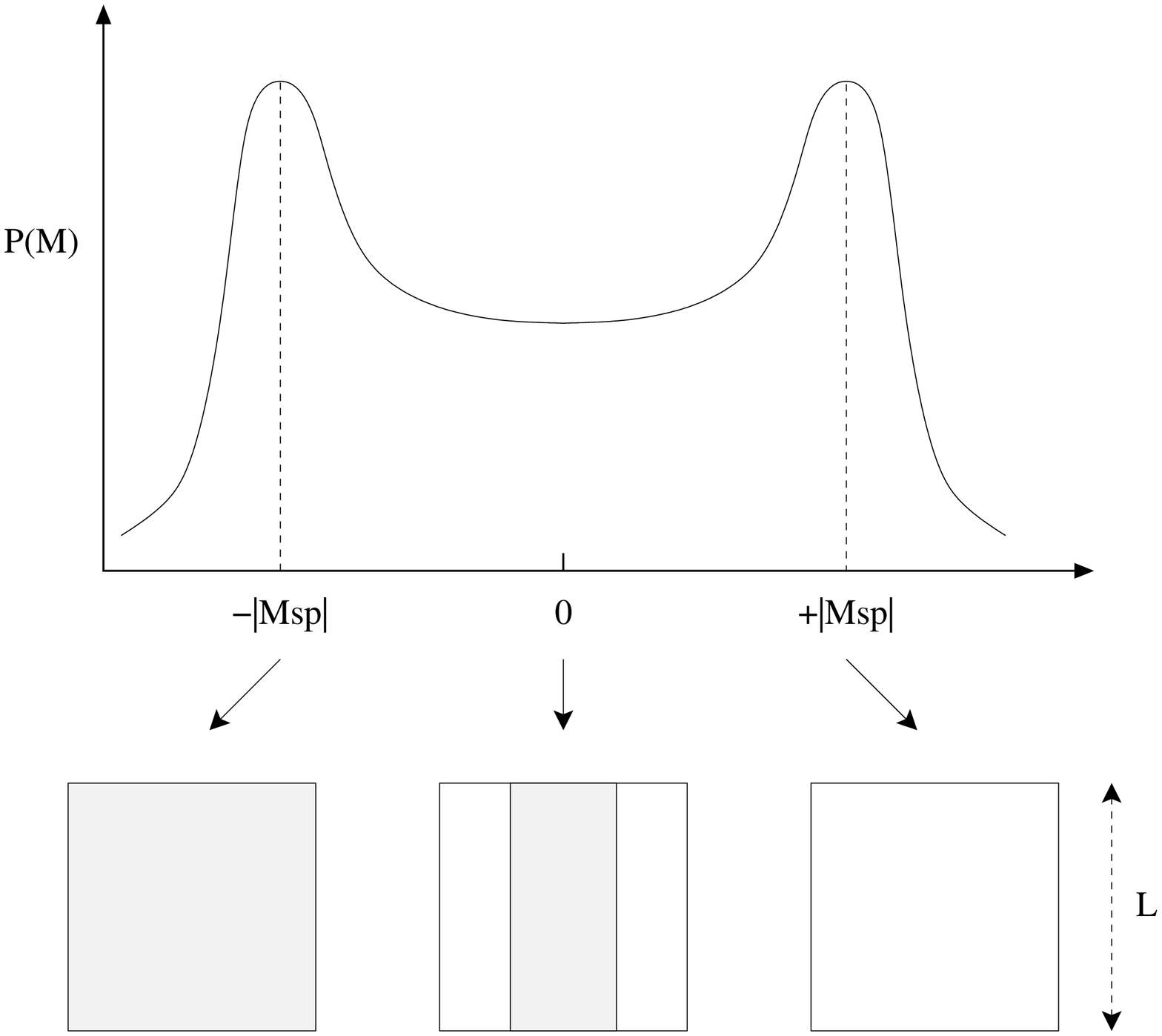,height=10cm,width=12cm,angle=0,clip=,bbllx=0pt,bblly=0pt,bburx=650pt,bbury=550pt}
  \caption{A schematic illustration of the probability density of
    the magnetization and how the representative spin distribution would populate a square latice of size $L$. The darker and lighter regions depict negative and positive spin repsectively.}\label{fig:sponmag2}
\end{figure}

Figure \ref{fig:sponmag2} schematically depicts this fact. The valley
shown linking the two peaks of the probability will thus be dropped
for lower temperatures and bigger lattice configurations. It should be
noted that even though the probability may be less it does always
exist and this has to be accounted for in data collection or
it may corrupt the results. We expect
the configuration to be relatively stable at the peaks but if
its magnetization has slipped down (fluctuated) to the center of the valley 
then it has an equal probability of climbing up either side of the
peaks, this is the crux of the spontaneous flipping. This aspect of
symmetry proves to also be the seed for a possible solution to this
problem. 

As an example of what has just been mentioned we note from 
Figure~\ref{fig:sponmag} where a fluctuation occurs just before $5000$
mcs and the magnetization peaks at $0$ from $-1$. The configuration is 
now in the middle of the valley and happens to go back to its previous 
state. The same phenomenon occurs just after $5000$ mcs but in this
instance chooses to flip to an opposite, but equally probable,
magnetization, from -1 to 1. 

If we now were to think of the implications of this spontaneous
flipping we come to the realization that it would cause an
averaging out of the mean magnetization, $\langle M \rangle$. This of
course has a detrimental effect on calculating the variance of the
magnetization and thus the susceptibility. 

\begin{figure}[htb]
  \centering
  \epsfig{file=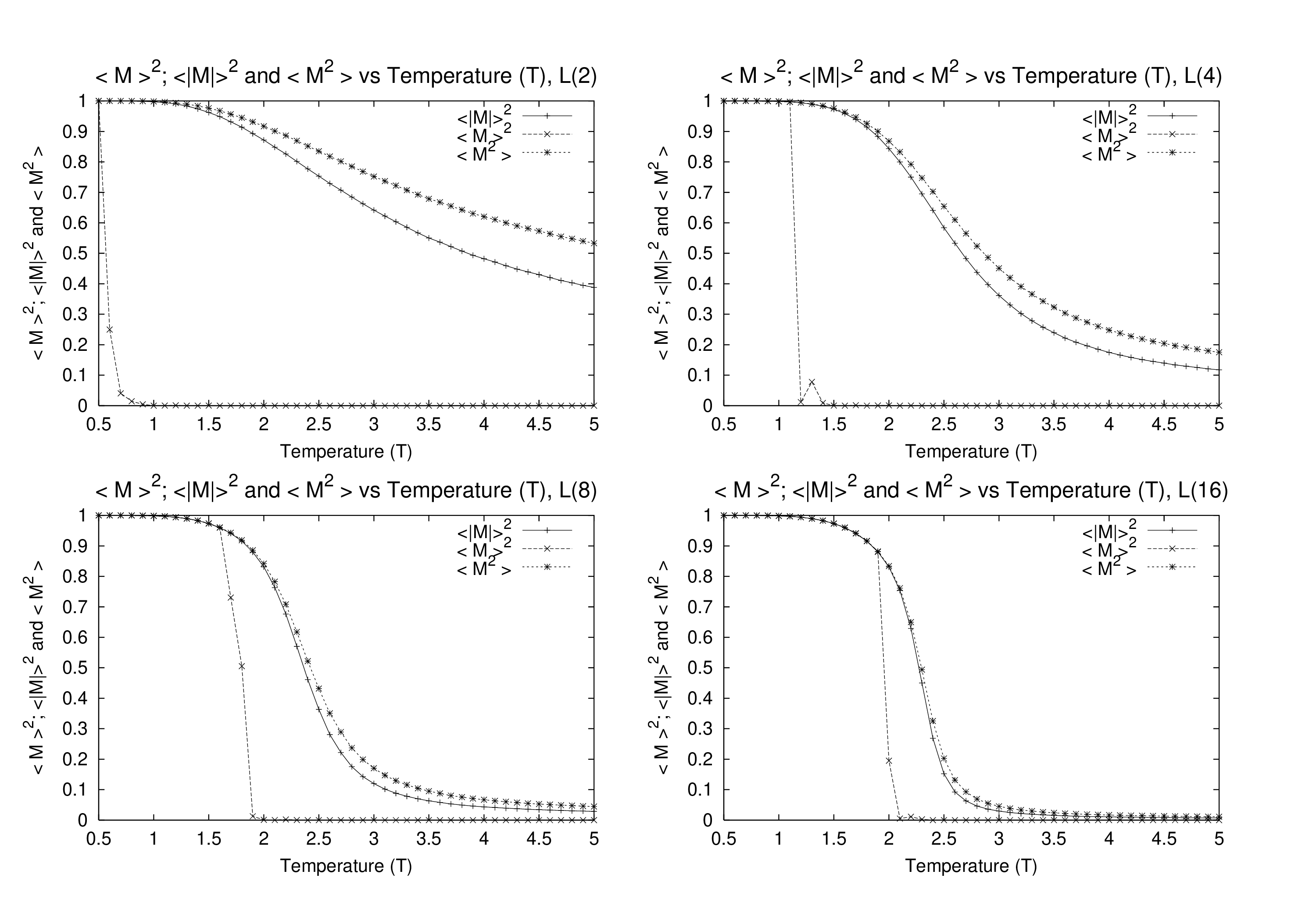,angle=270,width=\linewidth}
  \caption{This is an illustration of the differences in the normalized
    values of 
    $\langle M \rangle ^2; \langle |M| \rangle ^2$ and $\langle M^2
    \rangle$ with respect to temperature. The varying lattice sizes considerd are $2x2$ (top left); $4x4$ top right; $8x8$ (bottom left) and $16x16$ (bottom right).}
  \label{fig:variance_spon_mag} 
\end{figure}

This can be illustrated in
the Figure \ref{fig:variance_spon_mag} where the plot shows that
$\langle M \rangle^2$ remains zero for an extended period at low 
temperatures. This would cause the variance to peak at lower
temperatures. As the lattice size increases the spontaneous
magnetization is less likely to occur and the critical point moves
progressively to higher temperatures, this implies that the peak for
the susceptibility would approach the Curie temperature from the left
(lower temperatures). This is inconsistent with what the theory
prescribes \cite{Cardy}. We can also conclude that the greater
the number of  mcs we use the more likely we are to introduce spontaneous
flipping and thus averaging out of the mean magnetization which would
move the peak of the susceptibility more to the left.

\begin{figure}[hbt]
  \centering
  \epsfig{file=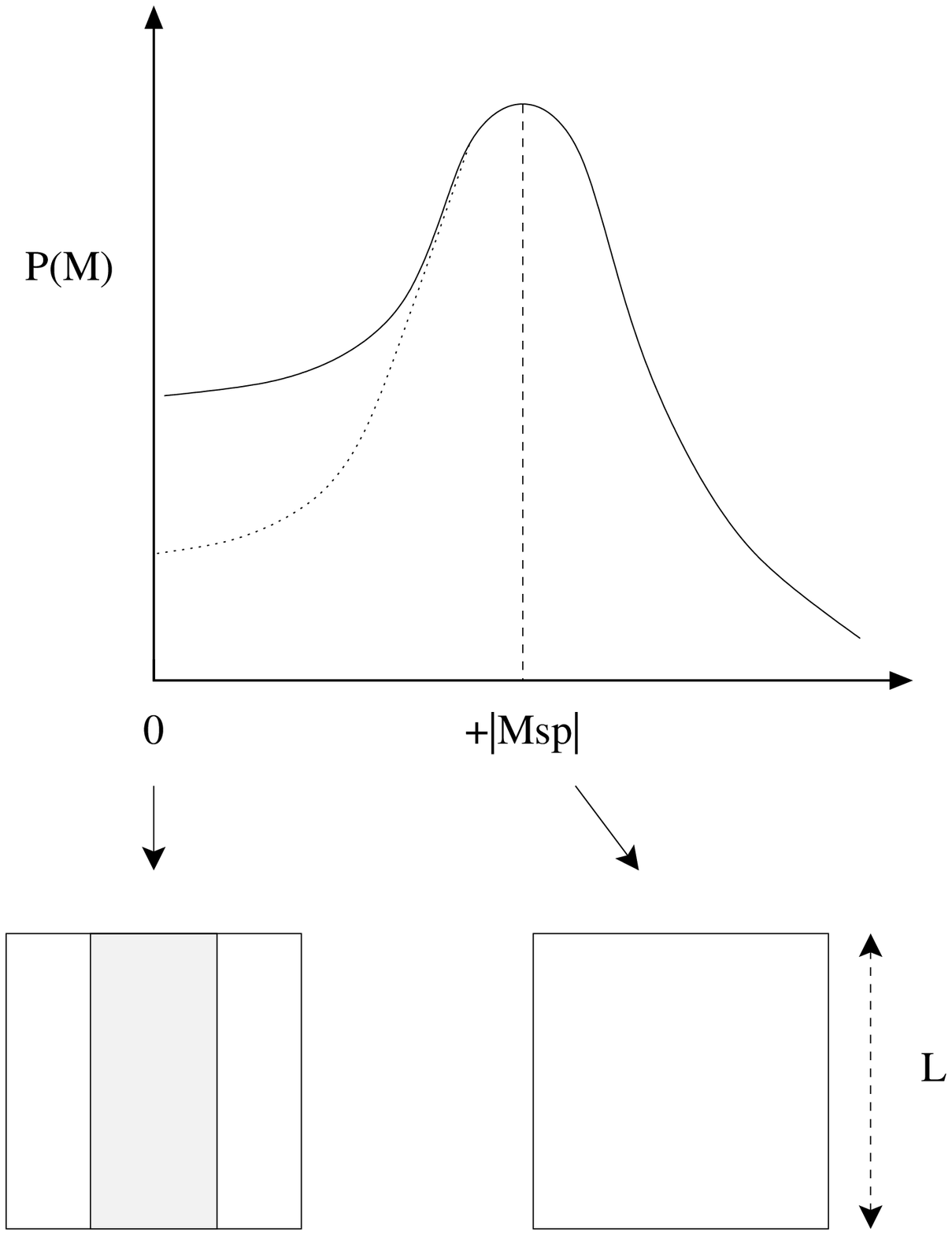,height=7cm,width=10cm,angle=0,clip=,bbllx=0pt,bblly=205pt,bburx=500pt,bbury=550pt}
  \caption{The solid line shows the revised probability density when using the absolute
    magnetization as opposed to the dotted line which represents the orginal propability density for magnetization.}\label{fig:sponmag3} 
\end{figure}

The solution to this problem lies in a characteristic that is at the heart 
of the problem, namely that there is an equal probability for the
magnetization to change to an opposite configuration or go back to its
previous configuration. Thus if we were to use the absolute
magnetization we would effectively be considering only one probability
peak (positive one) and the complications of averaging out the mean magnetization
would be overcome. Figure \ref{fig:sponmag2} thus changes to a
distribution shown in Figure \ref{fig:sponmag3} if we where to use the 
absolute magnetization. 

This modification doesn't come without a cost. In this instance it can 
be seen that we have averted the problems at low temperatures but end
up with weaker statistics at high temperatures. This is apparent 
from the fact that previously we had frequent fluctuations, in Figure
\ref{fig:sponmag2}, between positive and negative magnetization at
high temperatures resulting in a zero mean magnetization. However we
have only positive values to consider for the averaging of the mean
magnetization producing a non zero average. This effect is reduced
slightly since the valley is raised at high temperatures resulting in the
magnetization having a high probability of being close to zero. Fortunately this nonzero
average for the magnetization at higher temperatures is
inconsequential since it doesn't influence the Curie temperature and
appears only in the region above it.

At lower temperatures the shape of the distribution changes, as
indicated by the dotted line in Figure \ref{fig:sponmag3}. Thus the
magnetization remains relatively stable in a homogeneous
configuration. This is exactly the behaviour we expect and produces
good results.

\begin{figure}[htb]
  \centering
  \epsfig{file=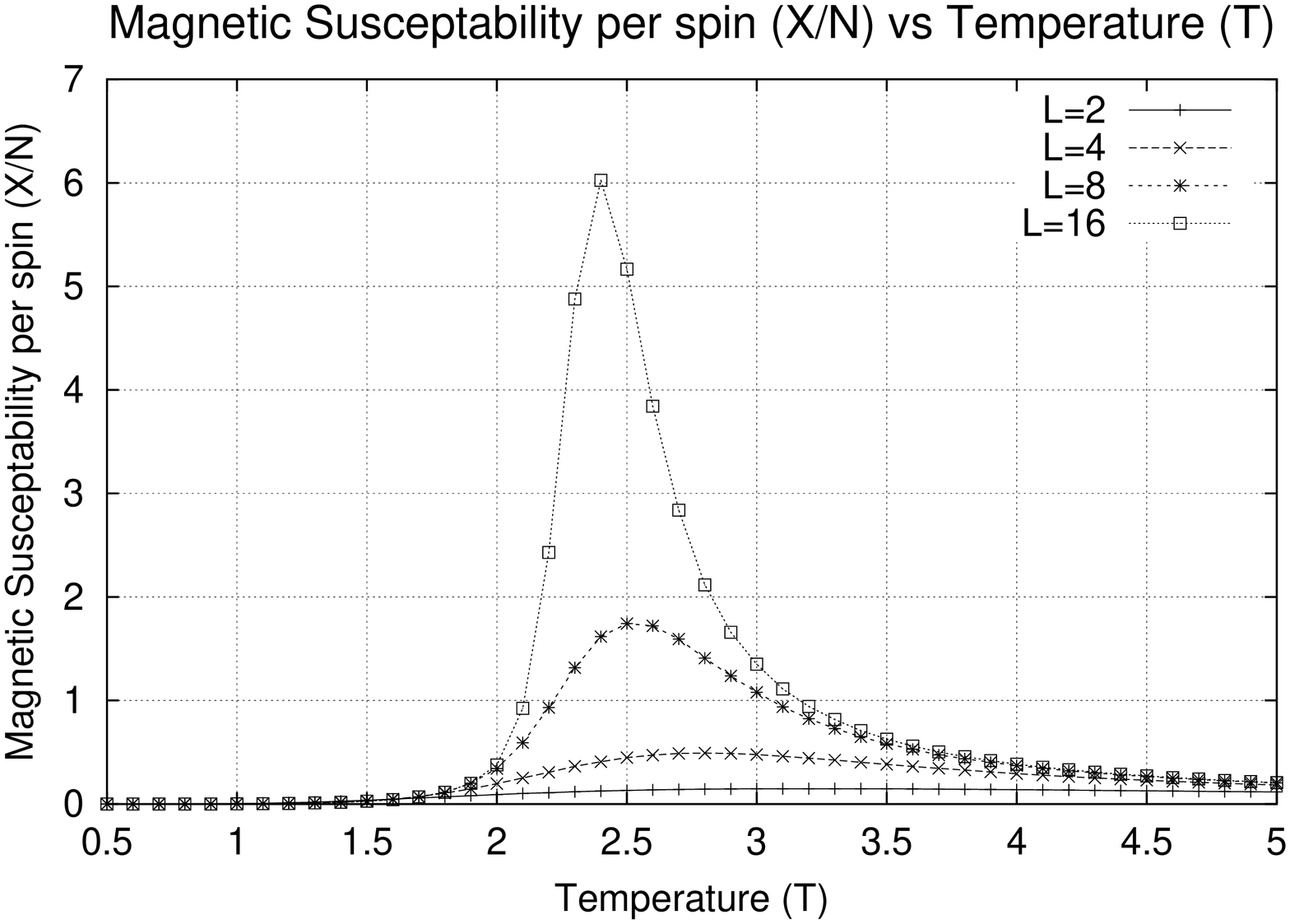,width=7cm,angle=270}
  \caption{This plot shows the differing results of the susceptibility
    for varying lattice
    sizes, $L\times L$.} \label{fig:X'vsT}
\end{figure}

\begin{figure}[htb]
  \centering
  \epsfig{file=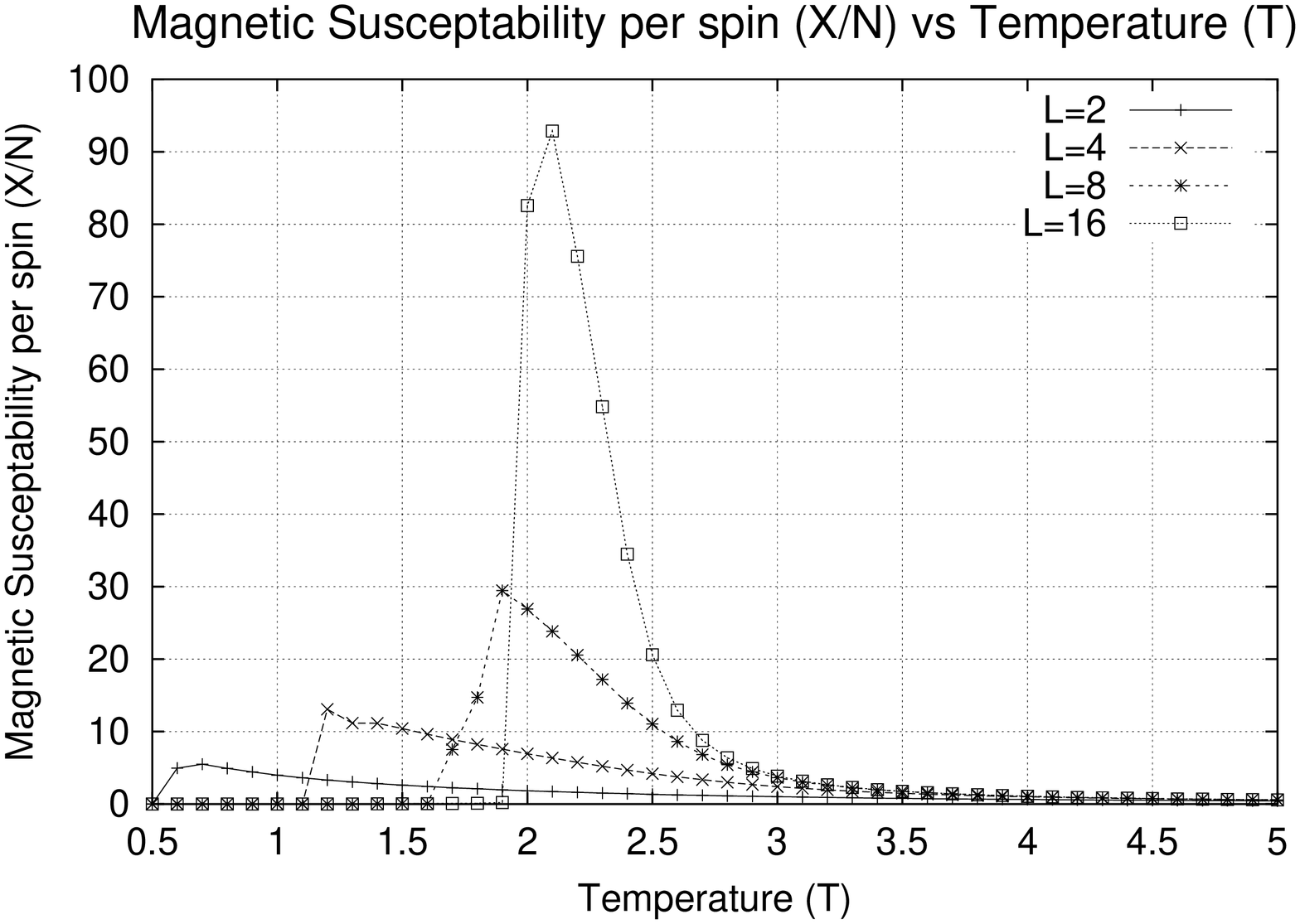,width=7cm,angle=270}
  \caption{This plot shows the differing results of the susceptibility
    for varying lattice
    sizes, $L\times L$.}\label{fig:XvsT}
\end{figure} 
The susceptibility that is produced in our data is thus not exactly
equivalent to the theoretical susceptibility, $\chi$, and we will be
distinguished as $\chi'$. The scaling characteristic of this
susceptibility is, however, equivalent to the theoretical value and only
varies by a constant factor above the Curie temperature.

\begin{equation}
  \label{eq:X'}
  \chi'= \frac{\langle M^2 \rangle - \langle |M| \rangle^2}{k_b T}
\end{equation}

A comparison can be made between $\chi'$ and $\chi$ in Figures
\ref{fig:X'vsT} and \ref{fig:XvsT} respectively. It is clear that a
marked difference in results occurs. This mistake becomes even more
evident if you were to use $\chi$ to get the critical exponent using
finite size scaling. Only $\chi'$ produces the correct finite size
scaling.

We now evaluate the plots produced using this technique discussed
thus far. The more distinctive character of the differing plots of Figure
\ref{fig:MvsT} produce more dramatic peaks in Figure
\ref{fig:X'vsT} of the magnetic susceptibility ($\chi'$) per spin versus
temperature as a result. This, once again, doesn't show an exact
divergence but shows a sharp peak for the $L=16$ lattice. This should be
strong evidence eluding to a second order phase transition.
\afterpage{\clearpage}


\subsection{Exact Results}

As pointed out in the beginning of this work there are significant challenges to
the calculation of exact solutions but we do need to evaluate the dependability of the
numerical process used in this discussion. For this purpose we restrict the
comparison between simulation and exact results to only a $2\times2$
This only serves as a good first order approximation but the corroboartion should imporve as the lattice size is increased.
 
The $2^{4}$ different configurations that can be listed can be reduced by
symmetric considerations to only four configurations.
\begin{figure}[htb]
  \centering 
  \epsfig{file=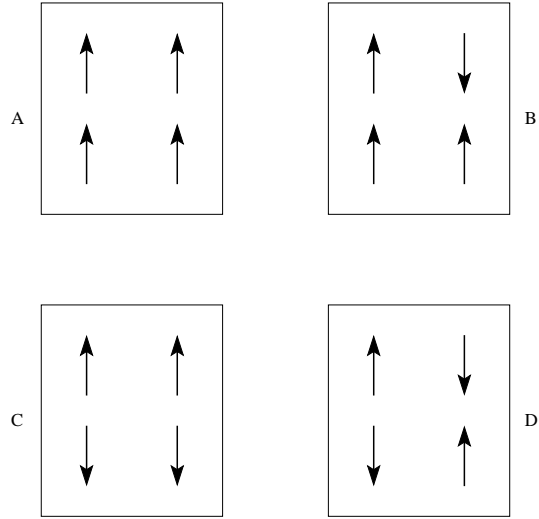,width=7.5cm}
  \caption[]{The four significant lattice configurations for a
    $2\times2$ lattice.}\label{fig:exactconfig}
\end{figure}
Configuration A is the fully aligned spin configuration and configuration 
B has one spin in the opposite direction to the other three while C
and D have two spins up and two spins down. To generate the full sixteen
different configurations we need only consider geometric variations of these
four configurations.

Using equations \eqref{eq:energy} and \eqref{eq:magnetization} we can
calculate the energy and magnetization for each of the four
configurations. Taking into account the degeneracy of each of these
configurations allows us to generate the exact equations for
calculating the relevant observables. These results are listed in
Table \ref{tab:exact}. 

\begin{table}[htb]
  \begin{tabularx}{\linewidth}{|Y|Y|Y|Y|}
    \hline
    Configuration&Degeneracy&Energy($E$)&Magnetization($M$) \\ 
    \hline
    A & 2 & -8 & +4,-4 \\ 
    B & 8 &  0 & +2,-2  \\ 
    C & 4 &  0 &0   \\ 
    D & 2 &  8 &0    \\
    \hline
  \end{tabularx}
  \caption{Energy and Magnetization for respective configurations illustrated in Figure \ref{fig:exactconfig}.} 
  \label{tab:exact}
\end{table}


Using equations \eqref{eq:prob_sum_M} and \eqref{eq:prob_sum_E} in
conjunction with the results of Table \ref{tab:exact} produces the
following equations 

\begin{equation}
  \label{eq:exct_partition_func}
  Z=2\ e^{8\beta J}+12+2\ e^{-8\beta} 
\end{equation}
\begin{equation}
  \label{eq:exct_E}
  \langle E \rangle = -\frac{1}{Z}[\ 2(8)\ e^{8\beta}+2(-8)\ e^{-8\beta}\ ]
\end{equation}
\begin{equation}
  \label{eq:exct_Esq}
  \langle E^2 \rangle = \frac{1}{Z} [\ 2(64)\ e^{8\beta}+2(64)\ e^{-8\beta}\ ]
\end{equation}
\begin{equation}
  \label{eq:exct_Mabs}
  \langle |M| \rangle = \frac{1}{Z}[\ 2(4)\ e^{8\beta}+8(2)\ ]
\end{equation}
\begin{equation}
  \label{eq:exct_Msq}
  \langle M^2 \rangle = \frac{1}{Z}[\ 2(16)\ e^{8\beta}+8(4)\ ].
\end{equation}

\begin{figure}[htb]
  \centering 
  \epsfig{file=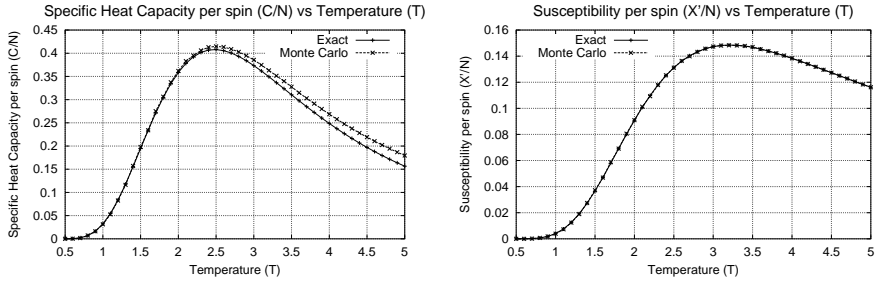,width=4.2cm,angle=270}
  \caption[hjsagh]{This plot shows a favourable comparison between the exact
    calculations done for the Heat Capacity per spin ($C/N$) and the
    Susceptibility per spin ($\chi'/N$)
    with the numerically generated solutions of the Monte Carlo simulation.}\label{fig:Exact}
\end{figure}

Applying equations
\eqref{eq:exct_E};~\eqref{eq:exct_Esq};~\eqref{eq:exct_Mabs} and
\eqref{eq:exct_Msq} to the formulas given in \eqref{eq:heat_capacity}
and \eqref{eq:X'} we can obtain the heat capacity and
susceptibility respectively. The comparison of these results are shown 
in Figure \ref{fig:Exact}. The results achieved by the Monte Carlo
method match the exact calculation exceptionally well and only deviate 
very slightly from the prescribed heat capacity at high temperatures.
 
\subsection{Finite size scaling}\label{sec:finite-size-scaling}

One of the limitations that the Ising Model confronts us with is the
finite size of our lattice. This results in a problem of recognizing
the specific point at which the phase transition occurs. This should be at a theoretical point of divergence but we are limited by the size of the lattice under consideration and thus dont see this divergence.
This effect is minimized by using periodic boundary conditions but
would only be resolved if we where to consider an infinitely sized
lattice as with the associated theoretical values for the phase
transition. It is thus necessary to use a construct
that will allow us to extrapolate the respective theoretical value
given the limited resource of a finite sized lattice. The aptly named
procedure of \emph{finite size scaling} is used to do just this.

It becomes useful to define a critical exponent to better understand
the nature of the divergence near the critical temperature. The
critical exponent, $\lambda$, is given by $\lambda
=\lim_{t\rightarrow0}\ \frac{\ \ln|F(t)|}{\ \ln|t|}$ or more commonly written
as $F(t)\sim |t|^\lambda$ where $t=(T-T_c)$. This exponent is important
in the sense that it offers a more universal characteristic for
differing data collected. This attribute will be taken advantage of
 and illustrated by showing that in a reduced unit plot the data
 collapses to a common curve with the same critical exponent. 

\begin{figure}[htb]
  \centering
  \epsfig{file=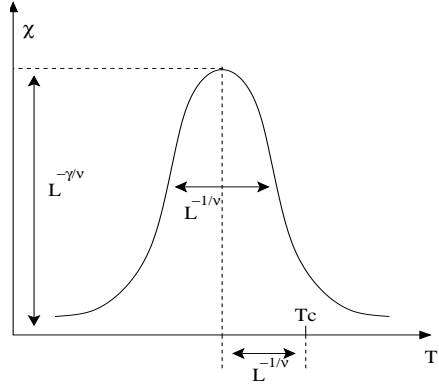,height=6cm,width=6cm,angle=0,clip=,bbllx=10pt,bblly=10pt,bburx=400pt,bbury=450pt}
  \caption{Theoretical illustration of Finite Size Scaling
    relationship for susceptibility $\chi$.}\label{fig:fss_show}
\end{figure}

The critical exponents relevant to the Ising model are as follows:
\begin{eqnarray*}
  \xi(T) \sim|T-T_c|^{-\nu} \\
  M(T) \sim(T_c-T)^\beta   \\
  C \sim|T-T_c|^{-\alpha} \\
  \chi \sim|T-T_c|^{-\gamma} \\
\end{eqnarray*}
If we examine the relationship between the lattice size, $L$, with respect
to the temperature relationship, $|T-T_c|$, we discover that
$|T-T_c|<<1$ as $L\rightarrow \infty$. Thus a critical exponent is
also applicable for the lattice size. This produces $L \sim
|T_c(L=\infty ) - T_c(L)|^{-\nu}$ which can in turn be used to reduce
the above exponents for the Ising model to a more appropriate form, in
terms of lattice size.

\begin{eqnarray}
  \xi(T) \sim|T-T_c|^{-\nu} \rightarrow L \label{eq:z-L} \\
  M(T) \sim(T_c-T)^\beta \rightarrow L^{-\beta/\nu} \label{eq:m-L} \\
  C(T) \sim|T-T_c|^{-\alpha}\rightarrow L^{\alpha /\nu} \label{eq:C-L} \\
  \chi \sim|T-T_c|^{-\gamma} \rightarrow L^{\gamma /\nu} \label{eq:X-L} 
\end{eqnarray}
This proportionality is illustrated in a schematic form in 
Figure \ref{fig:fss_show} of the relationship between the plot of a
certain observable and the respective critical exponent. In this
instance susceptibility was used as an example. 

We can now attempt to use equations \eqref{eq:z-L} - \eqref{eq:X-L} and
determine their appropriate exponents. This is simply done by taking
the peak values for the collected data of the observables and plotting
a $\ln$-$\ln$ graph that should yield a straight line with the gradient
being equal to the respective critical exponents. This procedure is
made easier since $\nu$ is equal to 1 for a two dimensional
lattice. An additional calculation for the observables of a $L=32$
lattice were done, since this would increase the number of data
points. This would also offset the poor statistics associated with the 
$L=2$ lattice and ultimately allow for a more accurate result.
\begin{figure}[htb]
  \centering
  \epsfig{file=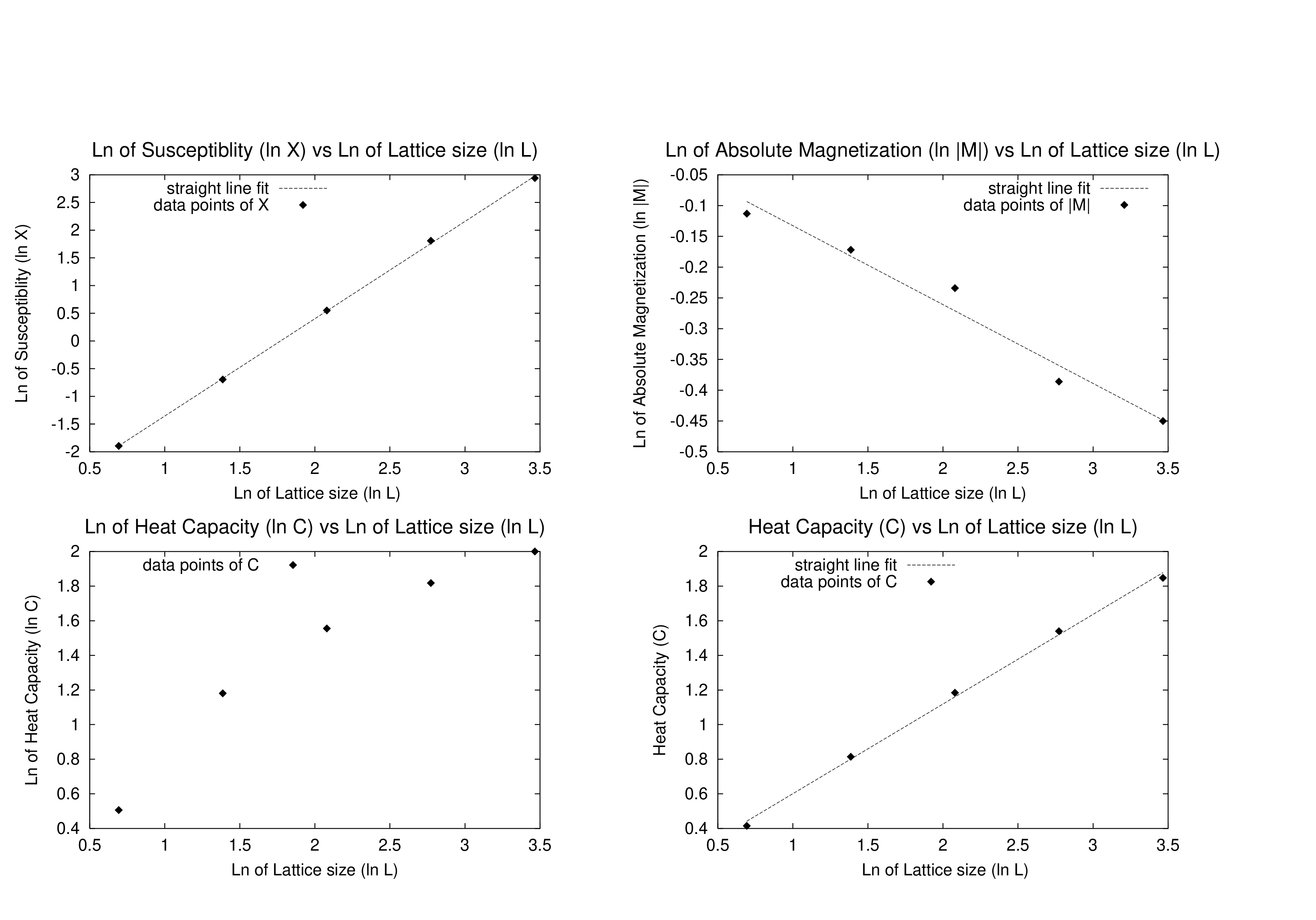,width=9cm,angle=270} 
  \caption{Plots obtained for the calculation of the critical
    exponents.}\label{fig:fss} 
\end{figure}



 
\begin{figure}[htb]
  \centering
  \epsfig{file=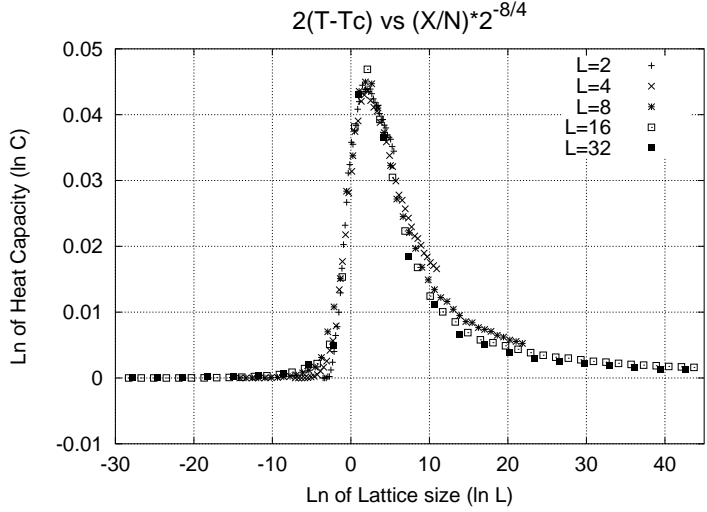,width=6.9cm,angle=270}
  \caption{This graph shows a reduced unit plot for $\chi'$ with a
    critical exponent of $\gamma=1.75$.}
\end{figure}
In Figure \ref{fig:fss} the calculation for the critical exponent is
shown. The answers are listed in Table \ref{tab:crit-exponent}. The graph for 
the heat capacity isn't a straight line and shows a curvature. The
reason is that $\alpha$ is zero in the two dimensional Ising model and 
should rather be interpreted as $C \sim \ C_0\ \ln \ L$. 
\begin{table}[htb]
\begin{tabularx}{\linewidth}{|Y|Y|Y|Y|Y|}
    \hline
    Quantity &
    Exponent & Finite Size Scaling (2d) &
    Theoretical (2d)\\ 
    \hline
    Magnetization &
    $\beta$  & 0.128 $\pm$ 0.008& 0.125\\ 
    Susceptibility  &
    $\gamma$ & 1.76 $\pm$ 0.01
    & 1.75\\ 
    Heat Capacity   &
    $C_0$     & 0.518 $\pm$ 0.02
    &0.500 \\ 
    \hline
  \end{tabularx}
  \caption{Displays the calculated and theoretical critical
    exponents.}
  \label{tab:crit-exponent}
\end{table}

We can in a naive sense attempt to determine the Curie temperature by
implementing the critical relation for the lattice size. This is
listed in Table \ref{tab:Tc-1}. It doesn't prove to be very accurate
and we thus have to implement a different notion in order to ascertain the 
critical temperature accurately. 
\begin{table}[htb]
  \begin{tabularx}{\linewidth}{|Y|Y|Y|Y|}
    \hline
    Lattice Size (L) & Estimated $T_c$ & Estimated $T_c(L\rightarrow\infty)$ &
    Theoretical  $T_c(L= \infty)$ \\ 
    \hline
    2 & 3.0  & 2.50 & 2.269\\ 
    4 & 2.8  & 2.55 & 2.269 \\ 
    8 & 2.5  & 2.43 & 2.269 \\ 
    16 & 2.4 & 2.34 & 2.269 \\
    32 & 2.3 & 2.27 & 2.269 \\
    \hline
  \end{tabularx}
  \caption{Listed critical temperatures calculated from the critical
    lattice size relation.} 
\label{tab:Tc-1}
\end{table}

The transition point can be determined by using the cumulant

\begin{equation}
  \label{eq:cum2}
  U_L=1-\frac{\langle M^4 \rangle_L}{3\langle M^2 \rangle_L}.
\end{equation}This calculation has to use double precision in order to retain a high
level of accuracy in calculating the critical point. To determine the
critical point we choose pairs of linear system sizes $(L,L')$. The
critical point is fixed at $U_L=U_{L'}$. Thus taking the ratio of
different cumulants for different sized lattices, $U_L/U_{L'}$, we will get an
intersection at a particular temperature. This is the desired critical
temperature. This procedure is not as straightforward as it may seem
and requires the cumulants to be collected very near to the transition 
point. Thus an iterative process may need to be employed
in order to narrow down the region of where the critical temperature is
located. 

\begin{figure}[htb]
  \centering
  \epsfig{file =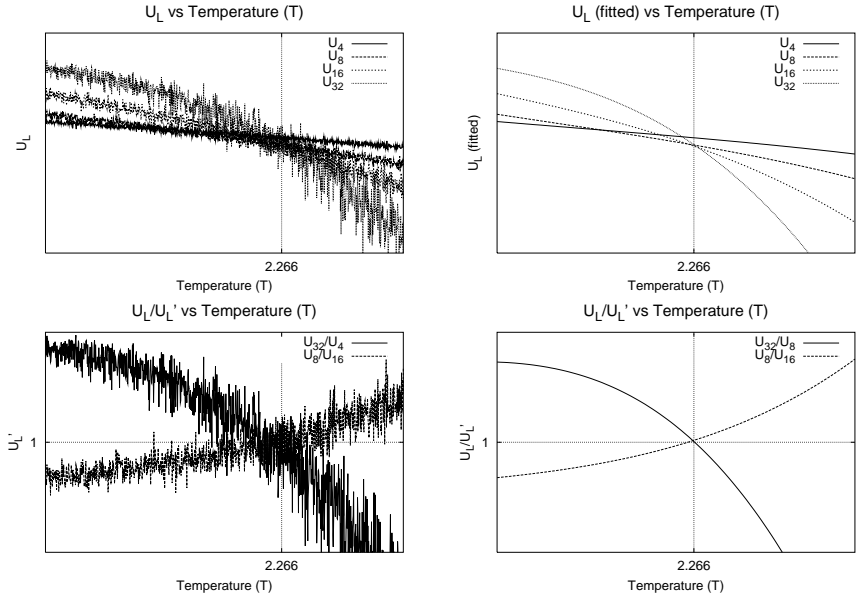,width=8.85cm,angle = 270}
  \caption{Plots of the cumulants ($U_L$).} \label{fig:cum}
\end{figure}

This analysis is done until an unique intersection is found
for all cumulants. This method is illustrated in Figure
\ref{fig:cum}. The $L=2$ lattice isn't shown since 
it doesn't exhibit the level of accuracy that we desire. From the
fitted graph of the cumulants it can be seen that intersection is
common to all the cumulants except for the $U_4$. This points towards
a poor value for this particular statistic and is thus not used, along
with $U_2$, in the ratios of the cumulants to determine the Curie temperatures.
The final analysis produces a result of a $T_c=2.266$ which agrees favourably with the
theoretical value of $T_c=2.269$. 
\subsection{Conclusion}
In this review we pointed out why an exact exposition of an Ising model for a ferromagnet is not easily achieved. An argument was proposed and motivated for making the computational problem far more tractable by considering a stochastic process in combination with the Metroplis-Hastings sampling algorithm. The numerical results produced by the Monte Carlo simulation compare
favourably with the theoretical results and are a
viable and efficient alternative to an exact calculation. The
requirements for producing accurate results are to 
consider large lattice sizes and a large number of Monte Carlo steps. The accuracy is very compelling even for small lattice sizes.

It is important to note and make provision for the potential of spontaneous magnetization to occur in a finite sized lattice. This can have serious consequences on the accuracy of the magnetization and the susceptibility which in turn will lead to incorrect results for the finite size scaling of these observables.
The occurrence and severity of spontaneous magnetization is directly proportional to the number of Monte Carlo steps used and inversely proportional to the lattice size considered. A practical means to overcome this complication is to use the absolute magnetization in the variance of the susceptibility instead of just the magnetization. This is an effective solution that produces good statistics only deviating slightly at high temperatures from the theoretical values. 

In conclusion a finite size scaling analysis was undertaken to determine the critical
exponents for the observables. These where in good agreement with the 
theoretical exponents. The critical temperature was also calculated using a
ratio of cumulants with differing lattice sizes and generated results which were in good agreement with the theoretical values. 
%
%
%
%

\nocite{Binder}
\nocite{Landau:76}
\nocite{Swendsen:87}
\nocite{Cardy}
\nocite{NumRecipe}
\nocite{Kinzel}
\nocite{Giordano}
\nocite{Hoffmann}
\nocite{Yeomans}
\nocite{Reichel}

\bibliographystyle{plain}
\bibliography{ising_arxiv}
\setlength{\oddsidemargin}{0cm}
\setlength{\evensidemargin}{0cm}
\setlength{\hoffset}{-0.5in}
\setlength{\voffset}{-0.5in}
\setlength{\marginparwidth}{0cm}
\setlength{\textwidth}{18.5cm}
\appendix
\section{Monte Carlo source code}
\begin{figure}[H]
  \epsfig{file=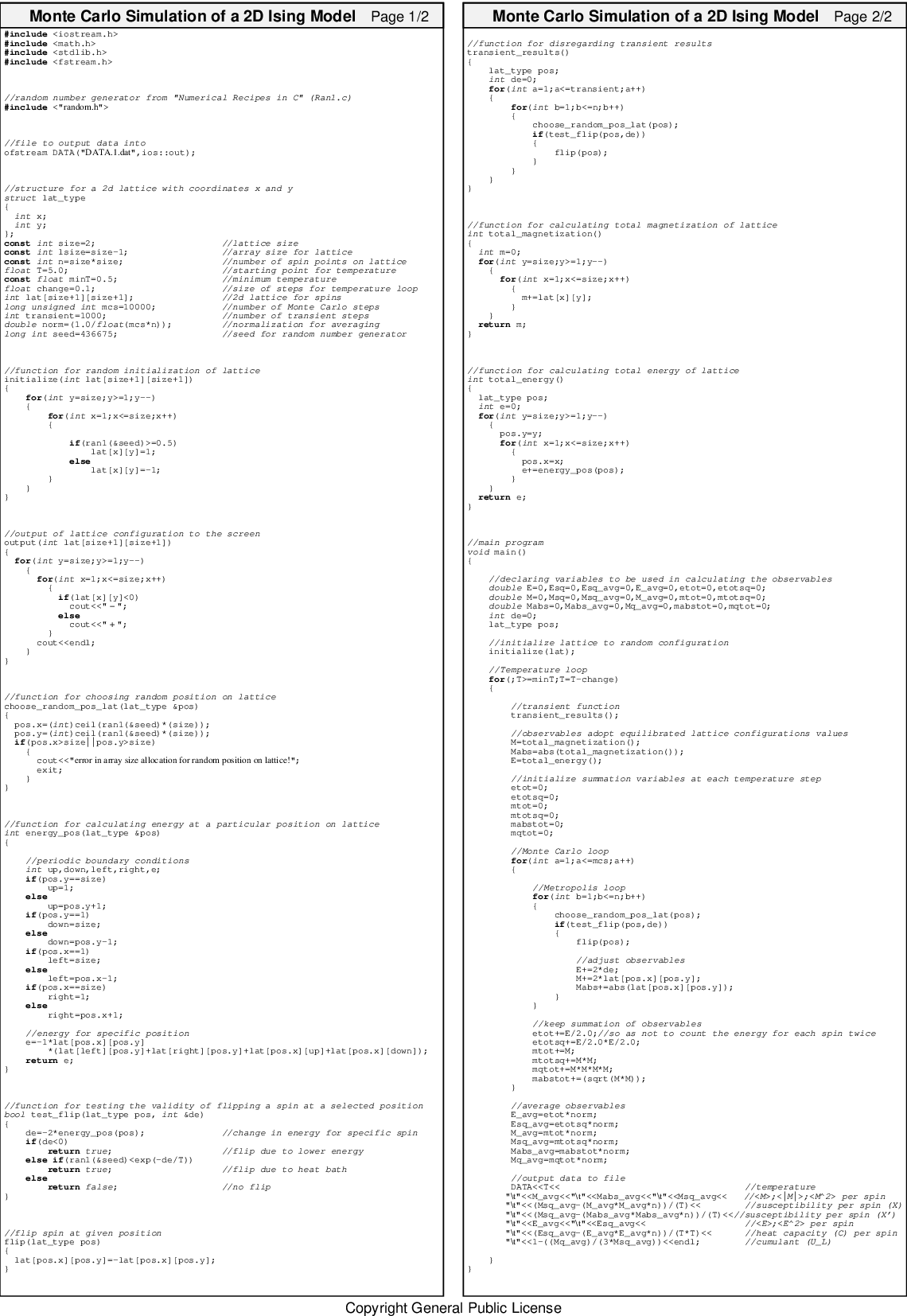,width=18.5cm,height=24cm}
\end{figure}

\end{document}